\documentclass[letterpaper, 10 pt, conference]{ieeeconf}  % Comment this line out if you need a4paper

\IEEEoverridecommandlockouts                              % This command is only needed if 
\usepackage{amsmath} % assumes amsmath package installed

\usepackage{graphicx}
\usepackage{multirow}
\usepackage{fancyvrb}

\title{\LARGE \bf
Gesture-based Human-robot Interaction for Field Programmable Autonomous Underwater Robots
}

\author{Pei Xu$^{1}$ % <-this % stops a space
\thanks{*This work was not supported by any organization}% <-this % stops a space
\thanks{$^{1}$Pei Xu is a student in the Department of Electrical and Computer Engineering,
        University of Minnesota, Twin Cities
        {\tt\small xuxx0884@umn.edu}}%
}

\begin{document}

\maketitle
\thispagestyle{empty}
\pagestyle{empty}

%%%%%%%%%%%%%%%%%%%%%%%%%%%%%%%%%%%%%%%%%%%%%%%%%%%%%%%%%%%%%%%%%%%%%%%%%%%%%%%%
\begin{abstract}

The uncertainty and variability of underwater environment propose the request to control underwater robots in real time and dynamically, especially in the scenarios where human and robots need to work collaboratively in the field. However, the underwater environment imposes harsh restrictions on the application of typical control and communication methods. Considering that gestures are a natural and efficient interactive way for human, we, utilizing convolution neural network, implement a real-time gesture-based recognition system, who can recognize 50 kinds of gestures from images captured by one normal monocular camera, and apply this recognition system in human and underwater robot interaction. We design A Flexible and Extendable Interaction Scheme (AFEIS) through which underwater robots can be programmed in situ underwater by human operators using customized gesture-based sign language. This paper elaborates the design of gesture recognition system and AFEIS, and presents our field trial results when applying this system and scheme on underwater robots.

\end{abstract}

%%%%%%%%%%%%%%%%%%%%%%%%%%%%%%%%%%%%%%%%%%%%%%%%%%%%%%%%%%%%%%%%%%%%%%%%%%%%%%%%
\section{INTRODUCTION}

On the land, keyboard, mouse, and other physical input devices are commonly used as a quite reliable mean to control computer and robots or interact with autonomous vehicles. Without physical input devices, the interaction between human and computer or robot still can be conducted smoothly through speech, infrared ray or other wireless communication approaches or/and sensors. However, in the case underwater, most wireless communication approaches and sensors based on electromagnetic signal would become useless due to the attenuation caused by the water, thereby imposing a huge challenge to interact with robots underwater. 

Usually, two methods are used for human and underwater robot interaction. For autonomous underwater vehicles (AUVs), we need to program the robot carefully before putting it in the water such that the robot can conduct its mission underwater independently without the intervention of operators. For remotely operated vehicles (ROVs), a tether is needed to connect the robot to a control platform on the land or on the surface of water such that the robot can be controlled by operators in real time on the platform. However, both of these interaction schemes have their limitations. For the first one, due to the lack of direct interaction between operators and robots, operators or programmers for AUVs have to consider all possible situations that a robot may encounter underwater and ensure the robot can properly deal with all kinds of accidents when programming the robot. This may be an impossible task due to the uncertainty of the underwater environment. In the interaction scheme for ROVs, tether may become a big problem if the underwater scene has complex topography; and due to that the ROV is controlled by operators on the remote control platform, some complex tasks that need operators to, according to the on-site situation, cooperate with robots in the field may be unable to be conducted.

In this context, we present A Flexible and Extendable Interaction Scheme (AFEIS) through which commands can be made by operators in situ to program and control AUVs in real time. Hand gestures are adopted as the input method for AFEIS in the case to interact with underwater robots. First, hand gestures are a natural way for most people to perform interaction and can be implemented conveniently without the support of external equipment. Besides, even underwater, information of gestures can be captured easily using optical cameras as long as the lighting condition is guaranteed. Although AFEIS is designed for interaction with underwater robots through hand gestures, it can be applied in other human-computer or human-robot interaction scenarios or/and through other interaction approaches like speech and picture-based signs or tags. 

Gesture recognition with high accuracy is a prerequisite to implementing such an interaction scheme. In normal application scenarios, gesture recognition using wearable electromagnetic devices and infrared ray can provide quite accurate recognition results. However, both of these methods are not employable underwater due to the confines of water. Traditionally, gesture recognition also can be implemented, based on images captured from optical cameras, by means of orientation histogram~\cite{ref:gr_hist}, hidden Markov model~\cite{ref:gr_markov}, particle filtering~\cite{ref:gr_pf}, and support vector machine~\cite{ref:gr_svm}. A common characteristic of these methods is that we must extract gesture features, like convexity defects, elongatedness and eccentricity, from images manually before feeding these features into classifiers. As for the final recognition result, it depends a lot on what features we take into account to describe gestures. Due to the lack in effective and comprehensive ways to describe various gestures based on certain manually extractable features, these methods usually only can provide satisfactory accuracy when recognizing very limited kinds of gestures. In this context, we use a convolution neural network (CNN) to perform gesture recognition and thus avoid extract features manually. The CNN we use has a simple structure such that the recognition system can run in real time on a platform with limited computation resources. We train a model which can recognize 50 kinds of gestures with over 99.6\% accuracy rate. When combining with AFEIS, a probabilistic model can be introduced to further improve the robustness of the whole system.

While introducing gestures as the interaction approach, AFEIS has two distinguishing characteristics: flexible and extendable. AFEIS allows operators to define the meaning of gestures by themselves. The `gestures' mentioned here are those used to control or program the robot in situ. They can be substituted by other control signals like speech and visual signs or tags in certain application scenarios. In AFEIS, the meaning represented by each gesture, instead of being fixed in the code, is parsed through independent configure files, which can be customized by each operator him or herself and which can be specified with respect to the hardware platform of the robot. Such a kind of design makes robots able to be controlled by a comfortable way accepted by each operator him or herself, and thus reduces the difficulty for operators to learn and use such an interaction system. Meanwhile, AFEIS decouples the input system, i.e. the hand gesture or other input or recognition system, and the control system, i.e. the system to directly control the action of robots. Therefore, it can be easily deployed on various robot platform with little additional work. Once AFEIS is deployed, we can through modifying configuration files make AFEIS support more control commands when the robot platform is extended. Furthermore, through dynamically loading different configure files during the process of interaction, operators can give various sets of control commands to the robot by a unique set of gestures, and thus the number of control commands that can be expressed by gestures increases significantly. Another key characteristic of AFEIS is to make robot `programmable` in situ. AFEIS, based on a set of simple syntax similar to programming language, parses signals obtained from the input systems. It does not command the robot according to each single gesture, but guides the robot to complete a series of commands according to a sequence of gestures posed by operators. Besides directly making commands to robots, it allows operators to define functions and set variables in situ when interacting with robots and thus makes the interaction process more flexible.

The rest part of this paper is organized as the following: a brief survey of related work and our comments are listed in Section \ref{sec:bg}; the implementation of AFEIS using gesture-based sign language for underwater robots are elaborated in Section \ref{sec:method}; the field test results are shown in Section \ref{sec:test}; and some improvement schemes that we are testing are presented in Section \ref{sec:conclusion}.

\section{BACKGROUND AND RELATED WORK}\label{sec:bg}

Gesture recognition is a fundamental link in the application of human and robot interaction, since hand gestures are chosen as the interaction way between human and robots in our design. Besides the recognition part, a potential problem involved in hand-based recognition system is hand detection and background removal. Without the device, like stereo cameras and infrared ray sensors to provide depth information, most of hand detection methods using monocular cameras are performed based on shape~\cite{ref:hand_shape}, color~\cite{ref:hand_color}, Harr features~\cite{ref:hand_harr} or context information~\cite{ref:hand_context} of hands. Due to the variousness and variability of hand gestures, the application of these methods usually requests some restrictions on the background that detectors are facing. None of hand detection methods using monocular cameras, according to what we have learned so far, can really work well in an arbitrary environment. This imposes a huge obstacle to the application of gesture recognition using monocular cameras in general cases. Fortunately, compared to the uncertain indoor or outdoor environment that the `on land' application may face, the environment underwater usually is much simpler with a relatively monotonous background. Moreover, the interaction between human and underwater robots usually happens in professional tasks. We can request operators to wear some special equipment, such as colored gloves, before performing the interaction. This is conducive to performing hand detection and facilitates the application of gesture recognition using monocular cameras in the underwater environment. 

In \cite{ref:artag_gesture}, the authors use ARTags to interact with robots underwater. In order to use this interaction method, operators must print ARTags and bind together into book form before going into the water. ARTags must be waterproofed in advance, and it is also a problem for operators to pick out the expected one from dozens of ARTags, which is unable to provide any intuitive meaning to readers.

In \cite{ref:motion_gesture}, the authors use the motion of the hand and arm instead of the hand gestures to perform
 interaction. A problem of this interaction scheme is that during interaction, operators have to hold the arm up and draw shapes by moving the hand and arm aloft. Such an interaction method is not so natural and it is unable for operators to draw shapes by the hand and arm precisely. The result provided by the authors only reaches about 97\% recognition rate with just five kinds of `gestures'. Another problem that may happen underwater but the authors do not address is that it sometimes is hard in the underwater environment to judge what causes the motion of the hand and arm. In the underwater environment, operators often cannot keep their poses very stable and the motion of hand and arm is often in fact caused by the movement of body but not that of the hand and arm itself. Besides, water flow often could push robots, especially robots with smaller size, to move slowly. This also could influence the judgment about the motion of hand and arm. In addition, the algorithm proposed by the authors is based on the analysis of point clouds, which leads to a higher demand in computation.

In \cite{ref:caddy}, the authors employ gloves with colored markers to implement gesture recognition underwater rather than directly basing the recognition on the whole hand or the glove covering the whole hand. The authors in the paper reveal neither the details of their gesture recognition algorithm nor the performance of their algorithm. However, a conceivable shortage is that the kinds of gestures that their algorithm can cover are limited since the recognition may be achievable only when there is enough space of colored markers shown in the camera.

Besides, in \cite{ref:caddy}, the authors define a set of syntax in the form similar to natural language as the way to translate gestures to commands accepted by robots. An effort to interact with robots using natural sentences expressed by gestures are made in the paper. However, in practice, we find that the syntax similar to programming language is more intuitive and acceptable by operators. Therefore, in AFEIS, we define a set of syntax similar to programming language but with simpler form. Besides commands the robot to make action through hand gestures, the syntax of AFEIS also supports function definition and variable setting, which are convenient for doing repetitive tasks.

\section{Methodology}\label{sec:method}

\subsection{Gesture Recognition}\label{sec:g_recog}
\begin{figure}[!t]
	\centering
	\twocolumn[{
		\includegraphics[width=\textwidth]{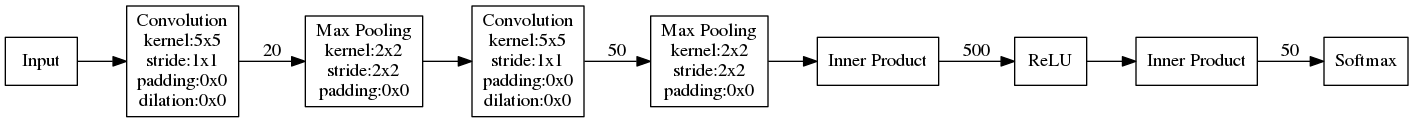}
		\caption{Structure of CNN classifier for gesture recognition. This network is obtained by modifying LeNet-5.}
	}]
	\label{fig:net_struct}
\end{figure}

\begin{figure}[!t]
	\centering
	\includegraphics[width=0.48\textwidth]{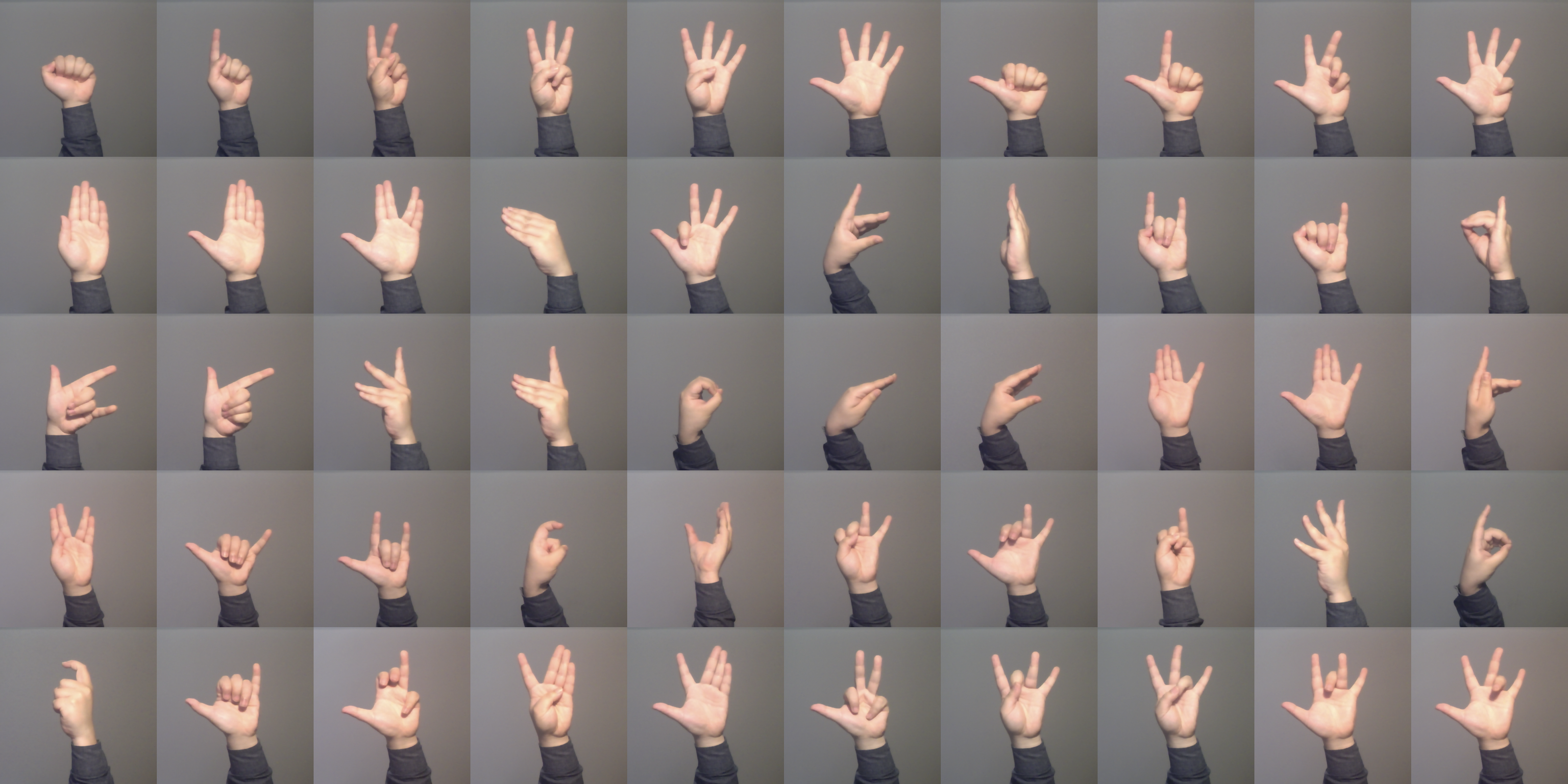}
	\caption{Thumbnails of all 50 kinds of gestures supported by the recognition model.}
	\label{fig:thumbnails}
\end{figure}

\begin{figure}[!t]
	\centering
	\includegraphics[width=0.48\textwidth]{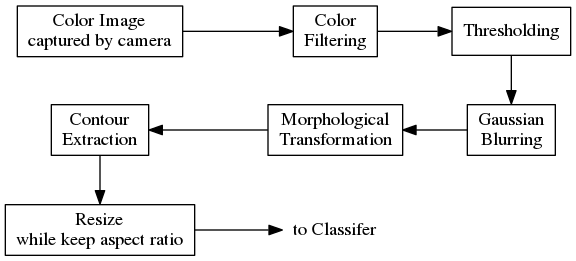}
	\caption{Flow of process to obtain the image fed into the CNN classifier.}
	\label{fig:process_flow}
\end{figure}

The CNN that we use to perform hand gesture recognition is obtained by modification to LeNet-5 \cite{ref:lenet}. The net structure is shown in Fig.~1. Differing from most CNNs that are popular at the present, the CNN in Fig.~1 performs recognition based on binary (black-and-white) images instead of color ones. Two concerns are taken into account to adopt binary images and the CNN with such a simple structure. First, for hand gesture recognition, color, even grayscale, usually is not a reliable feature. Most hand gestures are expressed by the combination of fingers posing at different positions in a quite small space. A subtle change in the lighting condition like the lighting angle and illumination intensity could significantly influence the looks of gesture images captured by monocular cameras and thus make the pre-trained model blind. The second concern is the computational intensity. Robots, especially those small ones, usually only can provide limited computation resources. We prefer a light-weighted recognition system, which can lead the whole system to running in real time at the cost of as few computation resources as possible. For this purpose, we also develop a light-weighted CNN framework named PNet to conduct gesture recognition. PNet is written in C++ for performance reason and designed as a light-weighted and flexible deep learning framework. The first version of PNet will be released later in 2017.

Our model supports 50 kinds of static, single-hand gestures, as shown in Fig.~\ref{fig:thumbnails}. It covers most static single-hand gestures that people can pose normally and is enough to satisfy the requirement of normal tasks in which operators have the need to interact with robots in the field underwater.

In our tests, hand detection is performed by color filtering. Operators are required wearing gloves with a certain color and the recognition system is calibrated according to the color before put into the robot. In the detection process, we only keep those pixels whose color is in a specific range and convert the image captured by the camera into a black-and-white one by thresholding. Then, after some augmentation, including Gaussian blurring and opening and closing morphological transformation, the binary image is cropped to keep only the largest contour region. Finally, the image is resized while keeping the aspect ratio before fed to the recognition system. The whole process of image processing is quite cheap, since the recognition system does not requests a piece of quite concise contour information such that there is no need for high-resolution image or complex technique to perform image processing.

\subsection{A Flexible and Extendable Interaction Scheme}\label{sec:afeis}

\begin{figure}[!t]
\begin{Verbatim}[fontsize=\scriptsize,commandchars=\\\{\}]
<explist>  ::= <exp> | <set-var> | <def-fn>
               | <change-keymap>
               
<def-fn>   ::= <def> <integer> \textbf{CMD_SEP} <cmdset> \textbf{END}
<def>      ::= \textbf{DEF} | \textbf{BEGIN}

<set-var>  ::= <set> <integer> \textbf{PARAM_SEP} <num> \textbf{END}
<change-keymap> ::=
               <set> <integer> \textbf{END}
<set>      ::= \textbf{SET} | \textbf{BEGIN}

<exp>      ::= \textbf{BEGIN} <integer> <do>
               <cmdset> [\textbf{CMD_SEP}] \textbf{END}
<do>       ::= \textbf{DO} | \textbf{BEGIN}
<cmdset>   ::= <cmd> [\textbf{CMD_SEP} <cmdset>]
<cmd>      ::= <function> | <call-fn>
               | <load-keymap> | <set-var>
               | <math-fn> | <exp> 

<function> ::= \textbf{FN} <arg-list>
<arg-list> ::= <arg> [\textbf{PARAM_SEP} <arg-list>]
<arg>      ::= <num> | <load-var> | \textbf{PARAM}
<load-var> ::= \textbf{CALL} <integer>

<call-fn>  ::= \textbf{CALL} <integer>

<load-keymap> ::= <set> <integer>

<set-var>  ::= <set> <integer> \textbf{PARAM_SEP} <num>

<math-fn>  ::= <math-op> <load-var> \textbf{PARAM_SEP}
               <math-arg-list>
<math-op>  ::= \textbf{+}|\textbf{-}|\textbf{*}|\textbf{/}
<math-arg-list> ::= <load-var> | <num>

<num>      ::= [neg-sign] <integer>
               [<decimal-point> <integer>]
<neg-sign> ::= \textbf{-}	
<decimal-point> ::= \textbf{.}
<integer>  ::= <digit> [integer]
<digit>    ::= \textbf{1}|\textbf{2}|\textbf{3}|\textbf{4}|\textbf{5}|\textbf{6}|\textbf{7}|\textbf{8}|\textbf{9}|\textbf{0}
\end{Verbatim}
\caption{AFEIS grammar defined by BNF. }
\label{fig:afeis_bnf}
\end{figure}

AFEIS is a scheme which is designed as a middleware between the recognition system and the robot's execution system. Fig.\ref{fig:afeis_bnf} shows the grammar of AFEIS parsing input signals, which are the recognition results obtained by the gesture recognition system in our tests. AFEIS itself does care neither the way to obtain input signals nor how the robot to execute each command. Its main task is to translate a series of input signals into a list of executable commands, based on which the robot can make actions sequentially. Besides, AFEIS provides the feature to define functions (\verb|<def-fn>|) and variables (\verb|<set-var>|) by operators in the field. It can record a series of commands as a custom-defined function, which is callable later (\verb|<call-fn>|) when needed; or store a number as a variable that can be modified by mathematical functions (\verb|<math-fn>|) or be loaded as a function argument later (\verb|<load-var>|). The \verb|<num>| in \verb|<exp>| after the \textbf{BEGIN} identifier indicates how many times the following \verb|<cmdset>| will be executed repeatedly, which is like a for-loop in common programming languages. In theory, operators can define any number of functions and variables in the field through AFEIS. The grammar of AFEIS is quite close to common programming language and is readily acceptable for most operators with whom we contacted. Through combining a function with custom-defined variables, AFEIS provides a way for operators to, in the field, define functions with arguments and dynamically call them.

AFEIS needs firstly to define some input signals to represent the basic symbols, \textbf{BEGIN}, \textbf{END}, \textbf{DEF}, \textbf{SET}, \textbf{DO}, \textbf{CMD\_SEP} (separator of commands), \textbf{PARAM\_SEP} (separator of parameters), \textbf{CALL}, 10 digits, decimal point and negative sign. Among those symbols, \textbf{DEF}, \textbf{SET} and \textbf{DO} can be replaced by \textbf{BEGIN}; and those representing \verb|<num>| can also be used as normal symbols for \textbf{PARAM}. Therefore, AFEIS only needs to occupy 5 slots of input signals but provide the features of function and variable definition and loop. Each of the rest available input signals can be used as a symbol representing a command \textbf{FN} or a parameter \textbf{PARAM}. Due to that when facing a \verb|<fn>|, AFEIS always parses the first input signal as the symbol of \textbf{FN} and the rest as \textbf{PARAM}, an input signal can, at the same time, have two definitions as symbols of \textbf{FN} and \textbf{PARAM} respectively. The \verb|<math-fn>| is optional and can be extended or disabled based on the request of tasks. Since most robots have the ability to perform mathematical operations, in practice, \verb|<math-fn>| also can be designed as a command accepted by the robot.

\begin{figure}[!t]
\begin{Verbatim}[fontsize=\scriptsize]
[system]         [system]
BEGIN=A          BEGIN=A
END=B            END=B
CALL=C           CALL=C
CMD_SEP=D        CMD_SEP=D 
PARAM_SEP=E      PARAM_SEP=E
DO=              DO=  ; =BEGIN by default
DEF=             DEF= ; =BEGIN by default
SET=             SET= ; =BEGIN by default

[fn]             [fn]
0=FORWARD        0=UP
1=LEFT           1=DOWN
2=RIGHT          2=SNAPSHOT

[param]          [param]
0=0              0=0 
1=1              1=1
2=2              2=2 
3=3              3=3
\end{Verbatim}
\caption{An example of two AFEIS configure files writing in INI format. The two files can be set simultaneously, and the operator can manually load either of them at a time when interacting with robots.}
\label{fig:afeis_keymap}
\end{figure}

\begin{figure}[!t]
\begin{Verbatim}[fontsize=\scriptsize,commandchars=\\\{\}]
A 1 D  // \textbf{DEF} 1       // define a funtion in slot 1
A 1 D  //   \textbf{SET} 1     //   load keymap 1  
1 1 D  //   \textbf{DOWN} 1    //   down 1 meter
2 D    //   \textbf{SNAPSHOT}  //   take a photo
A 0    //   \textbf{SET} 0     //   load default keymap 
B      // \textbf{END}

A 1 A  // \textbf{BEGIN} {3} \textbf{DO}  // do 3 times
C 1    //   \textbf{CALL} 1    //   call function in slot 1
B      // \textbf{END}         // execute
\end{Verbatim} 
\caption{An example of a signal sequence and its literal semantics to interact with a robot via AFEIS according to the configure files in Fig. \ref{fig:afeis_keymap}. The input signals are consisted of two parts. First, they define a function, which commands the robot to dive 1 meter and then to take a photo. Then this function is called 3 times and it is equal to command the robot to dive 3 meters and take a photo every time diving 1 meter.}
\label{fig:afeis_eg}
\end{figure}

\begin{figure}[!t]
\begin{Verbatim}[fontsize=\scriptsize,commandchars=\\\{\}]
(for i from 1 to 3 do
	(inform_robot, DOWN, 1)
	(inform_robot, SNAPSHOT)
)
(inform_robot, EXECUTE)
\end{Verbatim}
\caption{The general way through which AFEIS communicates with a robot after receiving the last \textbf{END} signal in Fig. \ref{fig:afeis_eg}.}
\label{fig:afeis_cmd}
\end{figure}

The definition of each input signal, in AFEIS, is defined in independent configure (\verb|keymap|) files instead of being hardcoded inside the code of AFEIS such that the recognition and execution systems are decoupled. Through such a way, the definition of each input signal can be set dynamically through loading different configure files during the interaction with robots. Besides, operators can define their own configure files and, by themselves, set the meaning of each input signal, i.e. the meaning of each gesture, thereby interacting with the robot by a most comfortable way accepted by themselves. This could largely reduce the difficulty of operators mastering the way to perform interaction with robots via AFEIS. Besides, due to decoupling, AFEIS can be easily deployed in most robot/computer systems with an interface to accept external commands.

Fig. \ref{fig:afeis_keymap} shows an example of two configure files writing in INI format. In the example, \verb|A|, \verb|B|, \verb|C| and \verb|0|, \verb|1|, \verb|2| and \verb|3| on the left side of equations are the indices of input signals each of which is corresponding to a hand gesture in our tests, while those on the right side of equations in the \verb|fn| and \verb|param| sections are the definitions of corresponding input signals when parsed as a symbol of \textbf{FN} and \textbf{PARAM} respectively. Additionally, there is neither the requirement that gestures and signals must be paired strictly nor the requirement that all gestures have to owe a definition. One signal can be represented by multiple gestures, while we can leave some gestures an empty definition if there is no need to use so many functions or parameters in our task.

Take the configure files in Fig.~\ref{fig:afeis_keymap} for example. If we want the robot to dive 3 meters and take a photo every time diving 1 meter, the operator can program the robot by giving the input signals shown in Fig. \ref{fig:afeis_eg}. After the last \textbf{END} signal is received, AFEIS will communicate with the robot and transfer commands to the robot by a way like what is shown in Fig. \ref{fig:afeis_cmd}. The robot simply stores all commands that it receives from AFEIS and executes the commands when meeting the command EXECUTE. The way shown in Fig. \ref{fig:afeis_cmd} is a general way through which AFEIS communicates with a robot. For robots who has an interface to parse a series of commands at once, the interaction between AFEIS and the robots can be done via just one communication.

\begin{figure}[!t]
	\centering
	\includegraphics[width=0.48\textwidth]{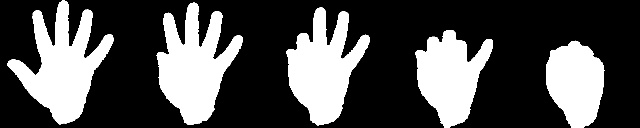}
	\caption{Some transient gestures are captured by the camera when the operator changes the gesture from `palm' to `fist'.}
	\label{fig:tran_gestures}
\end{figure}

Moreover, AFEIS also introduces a confirmation mechanism to improve the robustness of the gesture recognition system. When using a CNN classifier, any input instance will be classified as one of the classes that this classifier supports. This means that the gesture recognition system must make a mistake when facing any unrecognized gestures or noise, since it cannot tell us that the image it faces does not belong to any class it recognizes. One may require operators to try their best to accurately pose gestures. However, regardless of noise, due to the higher sampling speed, the camera always could capture some transient gestures when operators change their gesture into another one, as shown in Fig. \ref{fig:tran_gestures}. These transient gestures would lead the classifier to an unexpected recognition result. This means that in practice, it is unavoidable for the recognition system to encounter some unrecognized gestures and then make mistakes. In order to improve the robustness, AFEIS does not directly accept an input signal according to the recognition result obtained from a single image captured by the camera, but bases the signal on a probabilistic model which counts on the recognition results obtained in a time interval.

Suppose that $\mathcal{G}$ is a set of gestures recognized by the recognition system during a small time interval, i.e.
	\[
		\mathcal{G} = \{ g_{1}, g_{2}, \cdots, g_{n} \}
	\] 
where $g_{i}$ is the gesture recognized by the system according to the $i$-th image captured by the camera in the time interval. The discriminant function for AFEIS to accept the input signal as $s_{k}$ is defined as
	\[
		f_{k}(\mathcal{G}) = \Pr(s_{k} \vert \mathcal{G})
	\]	

A Markov model or Bayes risk estimator is employable to evaluate $f_{k}$. If we have no idea about the distribution of gestures with respect to a signal, a simple method is to count the occurrence of each gesture during a time interval, i.e.
	\[
		f_{k}(\mathcal{G}) = \frac{\sum_{i}r_{k, i}}{\vert \mathcal{G} \vert} \quad \text{if } \sum_{i}r_{k, i} > t
	\]
where $\vert \cdot \vert$ is the cardinality operation, $t$ is a threshold value and
	\[
		r_{k, i} = 
			\begin{cases}
				1, & \text{if } g_{i} \text{ is defined as signal } s_{k} \\
				0, & \text{otherwise}
			\end{cases}
	\]

Another key reason to introduce such a confirmation link in AFEIS is that human operators usually cannot perform operation synchronously with the sampling speed of the camera. Therefore, a buffer time is needed for operators to adjust their gestures when interacting with robots. Such a confirmation method employed by AFEIS will introduce a delay in the recognition process while improving the robustness. The length of the time interval used to collect $\mathcal{G}$ must be adjusted to balance the system's sensitivity and robustness.

Besides, when employing AFEIS, it is suggested to always leave some gesture definition empty for the sake of robustness. As what we talked above, when facing an unrecognized gesture or noise, the CNN classifier must make a mistake. However, if there are some gestures that have an empty definition, unrecognized gestures or noise has an opportunity to be labeled as those gestures not in use and then AFEIS will parse those false recognition result as nothing, thereby improving the robustness of the whole system.

\section{RESULTS}\label{sec:test}

\subsection{Gesture Recognition}

\begin{figure}[!t]
	\centering
	\includegraphics[width=0.48\textwidth]{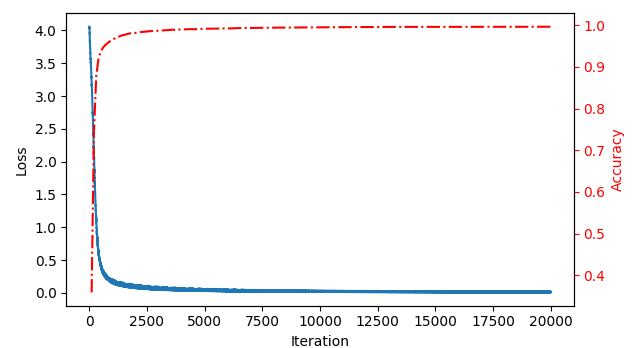}
	\caption{Loss and accuracy of our CNN model for hand gesture recognition during training and testing.}
	\label{fig:loss}
\end{figure}

We collect 75,000 static images of the right hand from 50 kinds of gestures as samples. For each gesture, 1,200 samples are used for training and 300 samples are collected for testing. Fig. \ref{fig:loss} shows the loss curve during training and the accuracy rate when applying the model to the test set. 

Each image captured by the camera is processed by the way that we talked in Section~\ref{sec:g_recog}. Finally, the obtained binary image is resized to 64-by-64 while keeping aspect ratio before fed to the CNN classifier.

\begin{figure}[!t]
	\centering
	\includegraphics[width=0.48\textwidth]{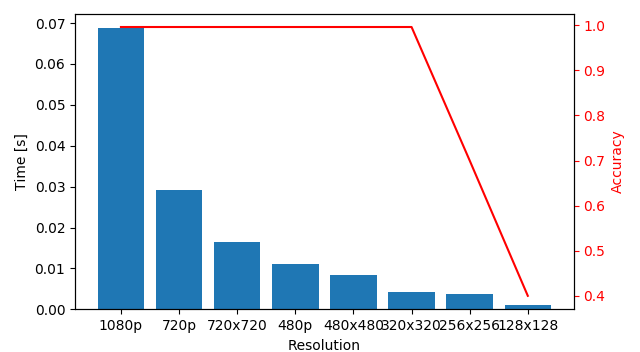}
	\caption{Time consumed by image processing and recognition accuracy with respect to different camera resolution.}
	\label{fig:resolution}
\end{figure}

OpenCV is employed to perform image processing, while PNet is used to run the gesture recognition model. No GPU is used during our practical tests. The consumed time from capturing image to obtaining a recognition result can be held within 0.02 second on a platform with DragonBoard 410c, which provides a CPU with 1.2GHz. This speed is achieved through limiting the camera resolution and thus dramatically reducing the time used for image processing. Fig. \ref{fig:resolution} shows the time consumed by image processing and recognition accuracy rate with respect to different camera resolution. A key reason to cause the low accuracy is that, considering the input size requested by the recognition model is 64-by-64, the contour region of gestures would be too small under lower resolution and thus cannot be extracted effectively.

During training, after 18,000 times of iteration, the model provides a satisfactory recognition result with an accuracy rate more than 99.6\% when applied on the test set. The error rate of each gesture is distributed averagely. This means that there is no significant difference in the accuracy for the classifier to recognize those gestures. Hence, in theory, operators can arbitrarily pick any ones in the 50 kinds of gestures through which to interact with robots.

\subsection{Field Trails}

\begin{figure}[!t]
	\centering
	\includegraphics[width=0.48\textwidth]{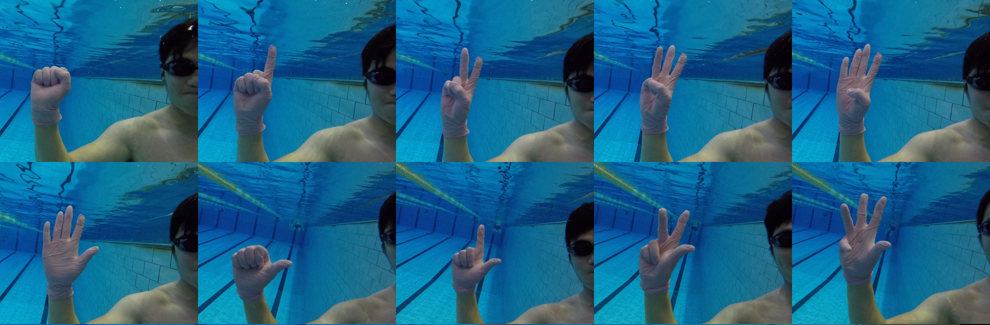}
	\caption{An exhibition of digits from 0 to 9 expressed by hand gestures during our swimming pool test.}
	\label{fig:trial1}
\end{figure}

\begin{figure}[!t]
	\centering
	\includegraphics[width=0.48\textwidth]{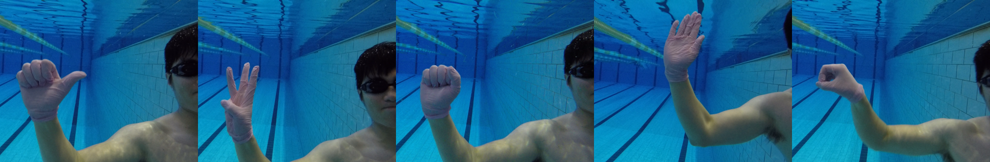}
	\caption{A sequence of gestures to command the robot to turn left 30 degrees and then take a picture. The definition of each gesture from left to right is Go Left, 3, 0, SEPARATOR, and SNAPSHOT.}
	\label{fig:trial2}
\end{figure}

A small ROV is used in our field trials for the sake of monitoring if the vehicle receives correct commands. Fig. \ref{fig:trial1} and \ref{fig:trial2} shows some pictures captured by the camera installed on the ROV during our swimming pool test. An interesting observation from Fig. \ref{fig:trial2} is that the operator was hardly able to keep his body or hand stable when interacting with the robot underwater, though he was tried to do so. This implies that the motion of hand usually is not a reliable signal during the interaction between human and robots underwater.

\begin{table}[!t]
\caption{Tasks}
\label{tb:tasks}
\begin{center}
\begin{tabular}{|c|p{2in}|c|}
\hline
Task & \multicolumn{1}{|c|}{Description} & Complexity$^a$\\
\hline
1 & Go Down 1 meter and Take a Photo. & 8\\
\hline
2 & Go Left 30 degree and Take a Photo. & 9\\
\hline
3 & Go to water Surface, Take a Photo and Go Back. & 9\\
\hline
4 & Swim Circle 3 times, Go Forward 2 meter, Take a Photo and Go Back. & 16\\
\hline
5 & Go Down 3 meter and Take a Photo every time going down 1 meter. & 8\\
\hline
6 & Go to Location 1$^b$, Take a Photo, Go to Location 2, Take a Photo and Go Back. & 13\\
\hline
7 & Follow Operator$^c$ and Take a Photo every 1 second. & 8\\
\hline
8 & Define a Function in the field to complete Task 5. & 14\\
\hline
\end{tabular}
\end{center}
\footnotesize{$^a$Complexity is the minimum number of input signals, i.e. the length of the sequence of gestures, needed to complete a task via AFEIS.\\$^b$ Location 1 and 2 are two pre-defined target destination coordinates.\\$^c$ Following action is achieved by tracking specific color.}
\end{table}

\begin{table}[!t]
\caption{Interaction Result associated with Empty Gesture Definitions}
\label{tb:task_acc}
\begin{center}
\begin{tabular}{|c|c|c|c||c|c|c|c|}
\hline
\multirow{2}{*}{Task} & \multicolumn{2}{|c|}{Empty} & \multirow{2}{*}{S/F$^c$} & \multirow{2}{*}{Task} & \multicolumn{2}{|c|}{Empty} & \multirow{2}{*}{S/F} \\
\cline{2-3}\cline{6-7}
 & \textbf{FN}$^a$ & \textbf{PARAM}$^b$&  &  & \textbf{FN} & \textbf{PARAM} & \\
\hline
\multirow{3}{*}{1}
 & 45 & 46 & 10/0 &
\multirow{3}{*}{5}
 & 45 & 44 & 10/0 \\
 & 10 & 10 & 10/0 &  & 10 & 10 & 10/0\\
 &  0 &  0 &  8/2 &  &  0 &  0 &  9/1\\
\hline
\multirow{3}{*}{2}
 & 44 & 46 & 10/0 &
\multirow{3}{*}{6}
 & 43 & 44 & 10/0 \\
 & 10 & 10 & 10/0 &  & 10 & 10 &  9/1\\
 &  0 &  0 &  9/1 &  &  0 &  0 &  7/3\\
\hline
\multirow{3}{*}{3}
 & 45 & 45 & 10/0 &
\multirow{3}{*}{7}
 & 44 & 45 & 10/0 \\
 & 10 & 10 & 10/0 &  & 10 & 10 & 10/0\\
 &  0 &  0 &  9/1 &  &  0 &  0 &  9/1\\
\hline
\multirow{3}{*}{4}
 & 45 & 46 & 10/0 &
\multirow{3}{*}{8}
 & 44 & 44 & 9/1 \\
 & 10 & 10 & 10/0 &  & 10 & 10 & 10/0\\
 &  0 &  0 &  7/3 &  &  0 &  0 &  7/3\\
\hline
\end{tabular}
\end{center}
\footnotesize{$^a$ \textbf{FN} means the number of gesture definition left empty for \textbf{FN} slots.\\$^b$ \textbf{PARAM} means the number of gesture definition left empty for \textbf{PARAM} slots.\\$^c$ S is for the number of successful interactions that reach the expected result, while F is for fail.}
\end{table}

Table \ref{tb:tasks} shows a part of tasks done in our field trials. An annotation is about Task 8. Basically, Task 8 does what we described in Fig. \ref{fig:afeis_eg}. It reaches the same operation goal with Task 5 but has a higher complexity. The complexity of Task 8 consists of two parts. The former part is about defining a function, while the other one is about calling the defined function in order to command the robot to make actions. The latter part only has a complexity of 6, which is lower than Task 5 and which usually would not increase when facing a more complicated task. That is to say that through allowing operators to define functions in the field, AFEIS provides an easy way for operators to command robots to finish repetitive complex tasks.

AFEIS allows a gesture having different definitions when it appears as \textbf{FN} and \textbf{PARAM} shown in Fig. \ref{fig:afeis_bnf} as well as allows gestures having empty definitions such that they would be parsed as nothing by AFEIS. To leave some gestures an empty definition is considered as a way to improve robustness, since some noise or unrecognized gestures would be classified as those having no definition and thus be ignored by AFEIS. In the field trials, we test the interaction performance in the case where there are different numbers of gestures that have an empty definition. Table \ref{tb:task_acc} shows partial results of our tests. In Table \ref{tb:task_acc}, the first row for each task is the case where we define as few gestures as possible; the second row is a general case where there are always ten gestures that are left empty; and the last row is the case where all gestures have their own definitions. As what we can see from Table \ref{tb:task_acc}, AFEIS performs poorly if no gesture has an empty definition. 

Another problem that we concern is the difficulty for operators to adapt the interaction way provided by AFEIS. It usually is not a problem for operators to remember the meaning of each gesture, since operators are allowed to define the meaning of each gesture by themselves. However, due to that on the ROV used in our tests, there is not anything like a screen that can provide a feedback to operators during the interaction, the operators have to do some practice beforehand in order to adapt the frequency in which AFEIS accepts input signals.

\section{CONCLUSIONS and FUTURE WORK}\label{sec:conclusion}

In this paper, we proposed an interaction scheme, named AFEIS, for human-robot interaction performed underwater in the field. This scheme uses a set of simple syntax similar to common programming language. It, besides providing a way for operators to directly make commands to robots, allows operators to define functions and set variables in the field and to call them later, thereby converting robots to programmable ones. We employ hand gestures as the way to perform interaction with robots underwater. A CNN model, supporting 50 kinds of static, single-hand gestures with more than 99.6\% recognition accuracy, is trained as the recognition system to provide AFEIS input signals such that AFEIS can control robots based on operators' gestures. AFEIS decouples the recognition system and the robot's execution system by using independent configure files to interpret input signals, and thus can be deployed in most robot systems with little additional work. By means of configure files, operators are allowed to, by themselves, define the input signal represented by each gesture such that the learning difficulty for operators to adapt AFEIS is quite low. In our field trials, AFEIS performs quite well and has the potential to be used in more hash underwater environment and for more complex tasks.

In the future work, we plan to further test AFEIS in the environment with a poor lighting condition and visibility. Besides, the CNN model that we trained is based on the right-hand images and we are collecting left-hand images in order to train a model for operators who prefer to use the left hand. We also have a consideration to allow AFEIS to work in a model that supports two-hand gestures. However, there are still some challenges when applying AFEIS for two-hand based interaction in the underwater environment.

\addtolength{\textheight}{-12cm}   % This command serves to balance the column lengths

%%%%%%%%%%%%%%%%%%%%%%%%%%%%%%%%%%%%%%%%%%%%%%%%%%%%%%%%%%%%%%%%%%%%%%%%%%%%%%%%

%%%%%%%%%%%%%%%%%%%%%%%%%%%%%%%%%%%%%%%%%%%%%%%%%%%%%%%%%%%%%%%%%%%%%%%%%%%%%%%%

%%%%%%%%%%%%%%%%%%%%%%%%%%%%%%%%%%%%%%%%%%%%%%%%%%%%%%%%%%%%%%%%%%%%%%%%%%%%%%%%

\end{document}